# MICROWAVE ABSORPTION IN BARIUM HEXAFERRITE NANOCOMPOSITES WITH RANDOM ANISOTROPY

*Jaume Calvo-de la Rosa[a,b],\*, Antoni García-Santiago[a,b], Joan Manel Hernàndez[a,b], Marc Vazquez-Aige[a], Jose Maria Lopez-Villegas[b,c], and Javier Tejada[a]*

[a]Departament de Física de la Matèria Condensada, Facultat de Física, Universitat de Barcelona, c. Martí i Franquès 1, 08028 Barcelona, Spain

[b]Institut de Nanociència i Nanotecnologia (IN2UB), Universitat de Barcelona, 08028 Barcelona, Spain

[c]Departament d'Enginyeria Electrònica i Biomèdica, Facultat de Física, Universitat de Barcelona, c. Martí i Franquès 1, 08028 Barcelona, Spain

\* jaumecalvo@ub.edu (corresponding author)

## Abstract

This work reports experimental evidence of random magnetic behavior observed in modified barium hexagonal ferrites. We observe a significant transition in the magnetic properties of this system when divalent cations ($Ni^{2+}$, $Cu^{2+}$, $Mn^{2+}$) incorporate into the structure and give rise to a magnetic nanocomposite. Such introduction randomly occurs throughout each sample and creates the conditions for such materials to behave as random anisotropy magnets. We verify the occurrence of such behavior in our samples by fitting the magnetization in approaching saturation to the corresponding theoretical model. We therefore analyze the microwave absorption capacities of random anisotropy magnets in the GHz range and predict large and broad absorption signals under certain conditions. The findings presented here postulate, for the first time, ceramic materials as promising random anisotropy magnets and underline their potential as microwave absorbers, in good agreement with recent theoretical models.

## 1. Introduction

Technological development relates closely to the search for materials with novel capabilities, especially for cutting-edge applications. Magnetic materials are present in many systems and devices and play a crucial role in their performance. In recent years, magnetic materials have been broadly investigated as microwave absorbers for electromagnetic shielding and stealth technology [1], [2], [3]. These technologies and related materials are attracting attention due, for instance, to their capacity to make vehicles invisible for radar detectors in military applications. Electromagnetic isolation of spaces or devices is another attractive use. More specifically, soft magnetic materials (SMMs) play a crucial role in applications that are looking for an increase in their operational frequency, such as electronics or telecommunications [4], [5]. Low remanent magnetization and coercivity, high magnetic susceptibility and permeability, and a small area of the magnetic hysteresis cycle characterize these materials. Therefore, the energy associated with their magnetization and demagnetization processes is meager, making SMMs excellent candidates for high-frequency applications. Some of the softest magnetic materials are known to be of amorphous or nanocrystalline nature, which are usually described as random anisotropy magnets [6].

The problem of random magnetic behavior has been widely investigated since the seminal work of Imry and Ma [7], mainly focusing on random anisotropy magnets (RAMs) [8], [9], [10], [11], [12], [13], [14], [15], [16], [17], [18], [19], [20], [21], [22], [23], [24]. Two competing effects characterize these systems: on the one hand, the magnetic moments tend to align along the directions of local anisotropy axes, which distribute randomly through the sample; on the other hand, neighboring magnetic moments interact via exchange coupling. So far, this behavior has been observed mainly in intermetallic materials containing rare earth elements, such as Gd-Fe-Ga [25], Gd-Er-Co [26], $HoFe_2$ [27], Fe-Ni-Pb-B [28], and Mn-B [28], [29]. More recently, Garanin and Chudnovsky [30], [31], [32], [33], [34] indicated that RAMs could act as powerful microwave absorbers in a wide frequency range and determined, both analytically and numerically, the power absorbed in several kinds of samples, for different cases of anisotropy and exchange interaction. Therefore, searching for materials that would lead to the realization of such a magnetic state is a field of primary interest.

Ferrites are one of the most widely used materials for magnetic applications. The possibility to modify ferrites, from a chemical and structural point of view, makes them very versatile. Their magnetic behavior may range from hard to soft according to their chemical and structural characteristics. A popular type is barium hexaferrite, which has a hexagonal crystalline structure and the chemical formula $BaFe_{12}O_{19}$. Different works demonstrated that these hexaferrites have appealing properties for high-frequency applications [35], [36], [37], [38], [39]. Several recent papers report how their behavior in the GHz frequency range changes when these structures are modified by adding certain divalent cations [40], [41], [42], [43]. These systems have been described as nanocomposites consisting of two phases: a hard magnetic one, corresponding to the barium hexaferrite, and a soft magnetic one, corresponding to the divalent ferrite. That prompted us to prepare and characterize these samples to test such systems for their application as microwave absorbers in the GHz frequency range.

As far as we know, this is the first work that reports the experimental observation of the appearance of random magnetic behavior in modified barium hexaferrites. To this effect, we describe the synthesis of hexaferrite-based pure and composite materials, followed by their chemical and structural characterization. Afterward, we analyze their magnetic properties, paying specific attention to their random magnetic behavior and stressing their impact on their properties and absorption capacity in the GHz range. The results make clear that modified barium hexaferrites can act as good microwave absorbers in a broad frequency range and suggest the possibility of designing ceramic RAMs, which are easier to prepare, chemically more stable, and less expensive than the traditional rare-earth-based intermetallic compounds.

## 2. Materials and methods

### 2.1. Materials

To synthesize the barium-hexaferrite powdered samples, we used the corresponding nitrates (*Scharlab S.L.*, as received) as reagents. On the one hand, we used iron and barium nitrates [$Fe(NO_3)_3 \cdot 9H_2O$ and $Ba(NO_3)_2$, respectively] for all samples. On the other hand, we used copper, nickel, and manganese nitrates [$Cu(NO_3)_2 \cdot 3H_2O$, $Ni(NO_3)_2 \cdot 6H_2O$, and $Mn(NO_3)_2 \cdot 4H_2O$] when we wanted to incorporate further cations to produce the composite structures (see Table I). In addition to reagents, we also used distilled water and sodium hydroxide (NaOH) pellets to synthesize the samples.

**Table I.** Identification of the four different powdered samples prepared.

| Sample ID | Description |
|---|---|
| HF | Pure barium hexaferrite |
| HF-Ni | Nickel-modified barium hexaferrite |
| HF-Cu | Copper-modified barium hexaferrite |
| HF-Mn | Manganese-modified barium hexaferrite |

### 2.2. Synthesis procedure

We prepared the different samples through a conventional co-precipitation approach under the experimental conditions reported elsewhere [44]. The chemical formula of sample HF is $BaFe_{12}O_{19}$, while we prepared the nanocomposites by adding a further cation M ($Ni^{2+}$, $Cu^{2+}$, or $Mn^{2+}$) in an atomic relationship defined by $BaFe_{12-x}M_{1.5x}O_{19}$, with $x = 2$.

We dissolved the corresponding reactants into distilled water in stoichiometric quantities and stirred at room temperature for 1 hour. Then, we added dropwise a 1 M basic solution of NaOH until pH = 10 to start the precipitation and stirred for one additional hour at 80 °C. Afterward, we dried it at 90 °C in an air atmosphere for 24 hours. We then ground the product in a mortar before thermal treatment in an air atmosphere at 1000 °C for 1 hour in an oven. We finally reground the product to prevent the powder from agglomerating.

### 2.3. Characterization

We analyzed the crystal structure and the chemical composition using X-ray diffraction (XRD), using a PANalytical X'Pert PRO MPD θ/θ Bragg−Brentano powder diffractometer of 240 mm radius, with Cu Kα radiation (λ = 1.5418 Å). We measured the particle size distributions by laser diffraction (LD) on a Beckman Coulter LS 13 320MW device. We performed the morphological analysis by scanning electron microscopy (SEM) using an ESEM Quanta 200 FEI, XTE 325/D8395 device.

We studied the magnetic properties by measuring first magnetization curves and full demagnetization curves at room temperature. First, we recorded the magnetization as the magnetic field increased to 2 T to ensure saturation. Then, we reduced the magnetic field to zero and reversed it in the opposite direction down to -2 T. We used a SQUID MPMS Quantum Design magnetometer to perform these experiments.

We measured the complex reflection and transmission parameters ($S_{11}$ and $S_{21}$, respectively) when we exposed the samples to microwaves in the GHz range. Figure 1 shows a schematic of the set-up we used to perform these measurements, using a Keysight E5071C ENA Series Vector Network Analyzer (VNA) and a custom-built sample holder made using 3.5 mm coaxial connectors. We placed a powdered sample in the central space of the holder, and we compacted it by pressing it to remove air and obtain a regular disk about 1 mm thick. Then, we connected the holder to the VNA through a two-port configuration to measure the reflected and transmitted signals.

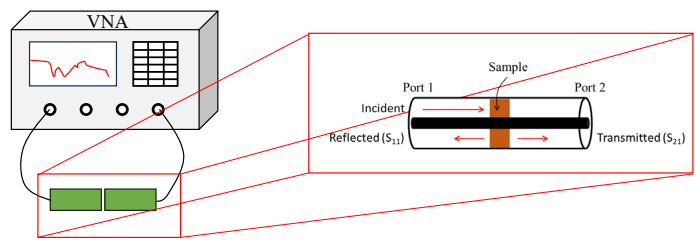

**Figure 1.** Schematic of the set-up used in the microwave absorption experiments. The right-side magnification shows a detail of the sample holder.

## 3. Results and discussion

### 3.1. Chemical and structural

We analyzed the four samples summarized in Table I by XRD to reveal their crystal structure and chemical composition. Figure 2 shows the experimentally obtained patterns.

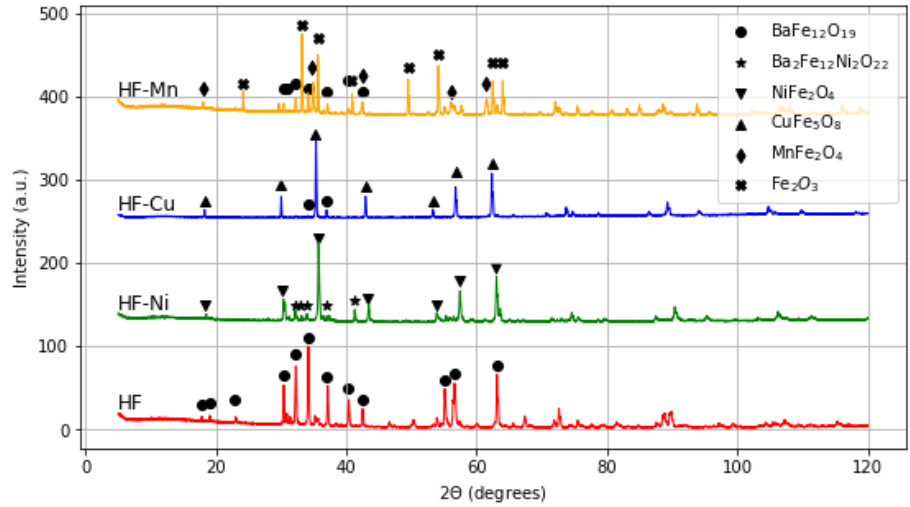

**Figure 2.** From bottom to top, experimental XRD patterns for the HF (red), HF-Ni (green), HF-Cu (blue), and HF-Mn (orange) samples. Symbols on top of the peaks identify the crystallographic phase corresponding to each diffraction (see legend for details).

The bottom red pattern corresponds to the HF sample. These data show that this sample consists of a pure hexagonal $BaFe_{12}O_{19}$ phase, as expected. On the other hand, the other three patterns correspond to hexaferrites that we modified by adding transition metal cations. In the case of the HF-Ni sample (green, second curve from bottom to top), the XRD analysis reveals the coexistence of a rhombohedral $Ba_2Ni_2Fe_{12}O_{22}$ phase and a $NiFe_2O_4$ phase. Even though a fraction of the $Ni^{2+}$ ions are incorporated into the $BaFe_{12}O_{19}$ and form a Ba-Ni ferrite, most of the $Ni^{2+}$ ions do not enter the structure of the hexaferrite and form the Ni ferrite phase. As for the HF-Cu sample (blue, third curve from bottom to top), we detect the presence of barium hexaferrite ($BaFe_{12}O_{19}$) together with copper ferrite ($CuFe_5O_8$). Finally, the HF-Mn sample (orange, topmost curve) consists of a mixture of $Fe_2O_3$, $MnFe_2O_4$, and $BaFe_{12}O_{19}$. Overall, we have obtained a sample made of pure barium hexaferrite and three types of composites based on the coexistence of a barium hexaferrite and the characteristic ferrite of the added cation, which is predominant.

Figure 3(a) shows the particle size distribution, expressed in terms of the percentage of volume occupied, as determined from the LD measurements performed in the four samples. We see a broad and multimodal distribution in all curves, with a tail ranging from a few to tens of micrometers, while the most significant particles (in volumetric terms) range between 200 and 400 μm, with slight changes among samples. Nonetheless, one must keep in mind that LD relates the diffraction of the laser beam with the particle size by assuming perfect spherical particles. Therefore, the particle shape might as well affect the width of the particle size distribution if the shape is non-perfectly spherical. According to the curves in this panel, the addition of the divalent cations leads, in general, to a reduction of the average particle size. Furthermore, if we express the particle size distribution in terms of the percentage of the number of particles having a determinate size [see Figure 3(b)], such distribution peaks instead at a value around 500 nm, indicating that most particles lie in this size range, with insubstantial variations from one sample to another.

To check and clarify these numerical results, we inspected the samples by SEM to confirm size distributions and examine their morphology. Figure 4 shows some of the images obtained with this technique. The low magnification images (left-side column, 1 mm scale) promptly reveal that there is indeed a broad distribution of particle sizes. This observation agrees well with the distributions reported in Figure 3. Moreover, the shapes appear irregular but still look spherical or spheroidal in some way, in the sense that we do not observe preferential directions nor cylindrical or ribbon-like shapes. Nevertheless, the particles are far from homogeneous when we change from one sample to another or look at different locations within each sample.

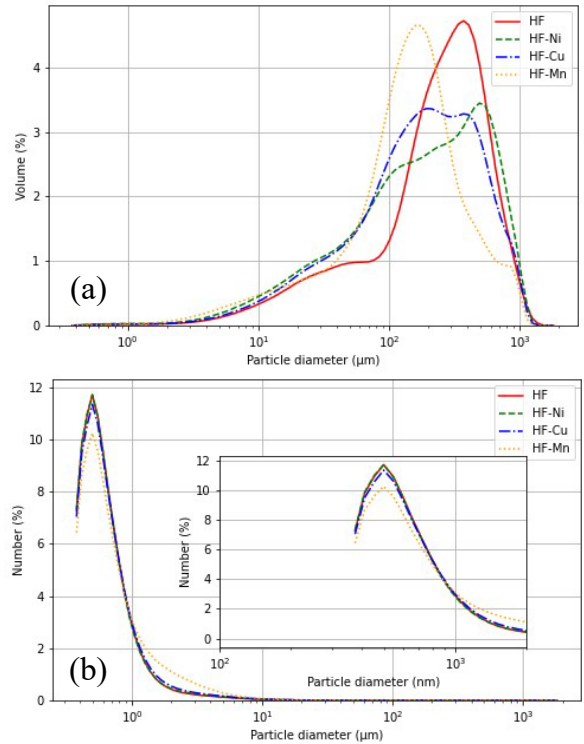

**Figure 3.** (a) Particle size distribution expressed in terms of the percentage of volume occupied. (b) Particle size distribution in terms of the percentage of particles having a certain size; the inset of the panel shows a magnification of the low-size range (hundreds to thousands of nanometers). In both panels, solid red stands for HF, dashed green for HF-Ni, dash-dotted blue for HF-Cu, and dotted orange for HF-Mn.

If we examine the high magnification images (right-side column, 30-µm scale), we may easily observe that the large particles form when much smaller ones agglomerate. These particles could have sintered during the thermal treatment at high temperatures and would lead to the formation of clusters of different sizes. Thus, we should assume that, instead of forming a wide range of particle sizes, the synthesis led to small particles that later agglomerated during the thermal treatment. Moreover, the shape of these particles is much more irregular than that of the larger clusters. In some cases, particles elongate along a preferential direction, while this effect does not occur in all cases. For instance, samples HF and HF-Ni [Figure 4(b) and (d)] are made of wire-like particles, while the small particles in samples HF-Cu and HF-Mn [Figure 4(f) and (h)] tend to be spherical. The shape of these particles also changes from one region to another in each sample. As mentioned before, this significant irregularity has a consequence on the LD measurements.

To finish this discussion, we must consider that the XRD patterns shown in Figure 2 confirm that all samples have an underlying crystalline structure that gives rise to the peaks at well-defined Bragg angles. The distinct width of the peaks reflects that X-rays are scattered by finite-size crystallites. We can thus apply the Scherrer equation, $L = 0.9\,\lambda/(\beta\cos\theta)$, to each relevant peak and estimate the size of the crystallites, $L$. In this equation, $\lambda$ is the wavelength of the radiation, and $\beta$ and $\theta$ are the full width at half maximum (FWHM) and the angular position of the peak, respectively [45], [46]. This calculation gives average crystallite size values of $43.5 \pm 3.3$ nm for HF, $39.0 \pm 4.5$ nm for HF-Ni, $43.6 \pm 10.6$ nm for HF-Cu, and $63.5 \pm 10.8$ nm for HF-Mn, confirming that each µm-size particle we identified in the SEM images consists of a collection of smaller crystallites of nanometer size. Accordingly, in the case of the modified samples, the particles shall be a heterogeneous, random mixture of crystallites of the original barium hexaferrite, as well as crystallites of the divalent cation ferrite.

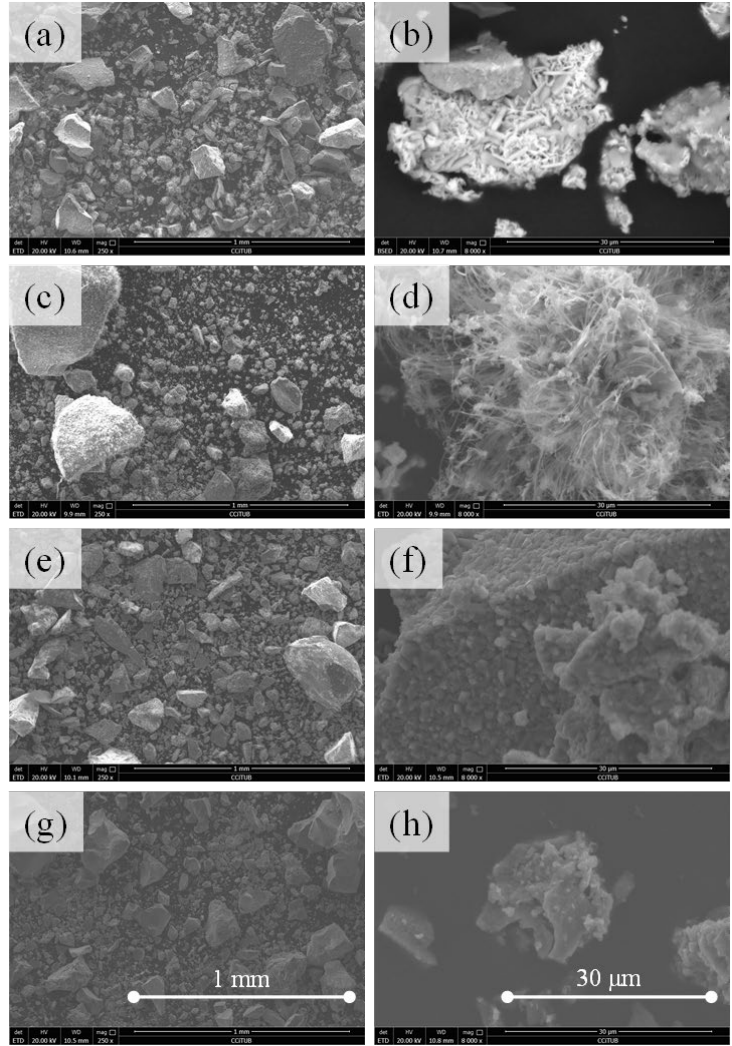

**Figure 4.** SEM images at x250 (left-side column, 1 mm scale) and x8000 (right-side column, 30 μm scale) magnification obtained for the HF (a, b), HF-Ni (c, d), HF-Cu (e, f), and HF-Mn (g, h) samples.

3.2. Magnetic properties

Now that we have a clear picture of the composition and structure of the samples, we move on to investigate their magnetic behavior. To this end, we performed an $M(H)$ measurement of each sample at room temperature. This process comprised the initial magnetization curve (which we acquired as the magnetic field evolved from 0 to 2 T) plus the demagnetization curve (which we recorded as we subsequently reduced the magnetic field from 2 T to –2 T). The main panels of Figure 5 show the results in the magnetic field range that goes from –2 T to 2 T. In addition, the inset in each panel presents a magnification of the two curves in a magnetic field range that may help discern between them. Table II contains the characteristic values extracted from the curves in this figure, such as the saturation magnetization ($M_S$), the remanent magnetization ($M_R$), the coercive field ($H_c$), and the magnetic susceptibility in the increasing part of the initial magnetization curve ($\chi$). Considering the reversibility of the demagnetization curve when the field reverses, we computed the area of the cycle as well.

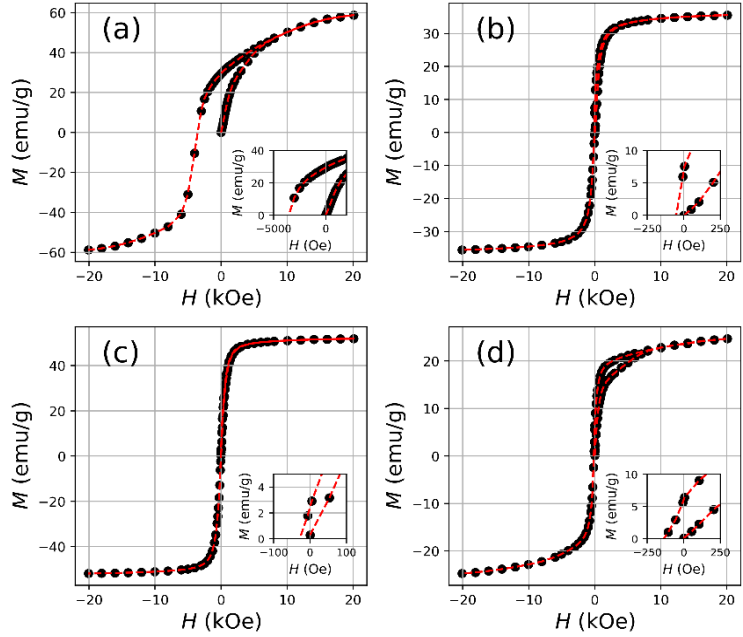

**Figure 5.** $M(H)$ measurements for the HF (a), HF-Ni (b), HF-Cu (c), and HF-Mn (d) samples. The black dots represent the experimental points, while the red lines correspond to the values interpolated by cubic splines to show the shapes of the cycles. Each panel contains an inset with a zoom to make visible the differences in reversibility between samples.

Figure 5(a) corresponds to the pure hexaferrite (HF), which has a low initial susceptibility [0.016 emu/(g Oe)] and a large saturation magnetization (58.8 emu/g), similar to well-known values found in the literature [47], [48]. We may also see that the cycle does not present a reversible behavior, and the remanence and coercivity are considerably large (29.5 emu/g and 3573.5 Oe, respectively). Conversely, the overall behavior in samples HF-Ni and HF-Cu [panels (b) and (c), respectively] is almost reversible, with high initial susceptibility [0.029 and 0.043 emu/(g Oe), severally], very low remanence (6.9 and 2.4 emu/g, respectively) and coercivity (49.8 Oe and 26.6 Oe, respectively). Nonetheless, the saturation magnetization values are similar to or slightly smaller than the value obtained for HF (35.8 and 51.9 emu/g, respectively). Finally, the addition of Mn [Figure 5(d)] results in a smaller saturation magnetization (up to 24.7 emu/g), a similar initial susceptibility [0.015 emu/(g Oe)], and a non-completely reversible behavior. Although there is some remanence (6.1 emu/g), the coercivity in this sample is low (137.2 Oe).

**Table II.** Characteristic values extracted from the $M(H)$ measurements. From left to right: saturation magnetization ($M_S$), remanent magnetization ($M_R$), coercive field ($H_c$), magnetic susceptibility in the increasing part of the initial magnetization curve ($\chi$), and the area of the cycle.

| Sample ID | $M_S$ (emu g$^{-1}$) | $M_R$ (emu g$^{-1}$) | $H_c$ (Oe) | $\chi$ (emu g$^{-1}$ Oe$^{-1}$) | Cycle area (kOe emu g$^{-1}$) |
|---|---|---|---|---|---|
| HF | 58.8 | 29.5 | 3573.5 | 0.016 | 477.0 |
| HF-Ni | 35.6 | 6.9 | 49.8 | 0.029 | 11.7 |
| HF-Cu | 51.9 | 2.4 | 26.6 | 0.043 | 2.0 |
| HF-Mn | 24.7 | 6.1 | 137.2 | 0.015 | 20.9 |

These data stress the effect the cation addition has on the barium hexaferrite. On the one hand, while the values of saturation in the modified samples are similar to or smaller than the value of the pure sample, there is a considerable increase in initial susceptibility, i.e., the material acquires a higher magnetic signal.

On the other hand, there is a remarkable drop in the coercivity and the area of the cycle, which suggests that the material becomes magnetically softer. Such transition to a very soft behavior, together with the nanometric nature of the crystalline structure of the modified samples, encouraged us to examine the possibility that they would behave as random anisotropy magnets.

Composites that incorporate barium hexaferrite and a second magnetic component have been described in the literature [42], [43], [49] as a set of two kinds of magnetic particles whose anisotropy axes randomly orient through the sample, and between which an exchange interaction takes place. This was precisely the basis for the theoretical description of random anisotropy magnets from the very beginning [7], [8], [10], [11], [19], [20], [23], [24]. Considering that we can describe the random anisotropy and the exchange interaction respectively by the fields $H_r$ and $H_{ex}$, the following law gives the magnetic field dependence of the magnetization in approaching saturation:

$$\delta M/M_0 = (1/30)(H_r/H_{ex})^2 (H_{ex}/H)^{1/2} \int_0^\infty dy C(y) y^2 exp(-y(H/H_{ex})^{1/2}) \quad (1)$$

In this equation, $M_0$ is the saturation magnetization, $\delta M \equiv M_0 - M$ is the change in magnetization in approaching saturation, and $C(y)$ is the function that correlates the orientations of the anisotropy axes in different locations through the sample in real space. If we approximate this expression at low and high magnetic fields, we obtain the following equations:

$$\delta M/M_0 = (v_c/30)(H_r/H_{ex})^2 (H_{ex}/H)^{\frac{1}{2}} \quad if \ H \ll H_{ex}, \quad (2)$$

where $v_c$ contains the correlation function $C(y)$ and roughly equals the unity, and

$$\delta M/M_0 = (1/15)(H_r/H)^2 \quad if \ H \gg H_{ex} \quad (3)$$

The $H^{-1/2}$ approach to saturation given by Equation (2) occurs at low fields in the so-called weak anisotropy case; that is, when the zero-field ferromagnetic correlation length throughout the material, $R_f \sim R_a (H_{ex}/H_r)^{1/2}$, exceeds the average size of the grains, $R_a$ [23]. In systems such as our samples, where crystallites, whose sizes are 40–60 nm (as from XRD measurements), play the role of grains, the scenario is the opposite. Consequently, given that the exchange interaction between grains is proportional to $1/R_a^2$ [24], such magnitude becomes irrelevant. Thus, the system lies in the so-called strong anisotropy case, in which it behaves as if it were composed of independent grains with randomly oriented anisotropy axes, and only such anisotropy determines the approach of magnetization to saturation. Therefore, $H^{-2}$ would govern the whole field dependence for all values at which $\delta M \ll M_0$ [23], [24]. Figure 6 presents the results obtained for all four samples by fitting the experimental data (blue dots) to Equation (3). We choose the cutoff fields individually for each sample to satisfy the condition $\delta M \ll M_0$.

Panel (a) in this figure shows that the relationship between $\delta M/M_0$ and $H^{-2}$ is not linear for the HF sample, indicating that the pure hexaferrite does not behave as a RAM. Even though the fitting line and the experimental values are numerically close, the experimental trend is not linear, crossing the two extremes of the fitting line and showing a sort of bent shape. This behavior changes when we add the divalent cations, and a good linear fit happens to occur for all three modified samples [see panels (b) to (d)]. In these cases, the error is one to three orders of magnitude lower than the one for HF (see the subplots in the figure). Therefore, due to the validation of the $H^{-2}$ dependence close to saturation, we can consider that the three modified samples behave as RAMs.

We extracted the anisotropy field ($H_r$) from the slope of the best linear regression of the experimental data with $H^{-2}$. The values obtained were 275.8 Oe for HF-Ni, 157.4 Oe for HF-Cu, and 573.8 Oe for HF-Mn. Moreover, by fitting the whole $C(y)$ function to the experimental data, we obtained identical results (not shown) when we used either a Gaussian or a decreasing exponential function. That agrees with the interpretation that our system consists of randomly oriented independent grains with a strong anisotropy dominance upon exchange interaction between grains [17].

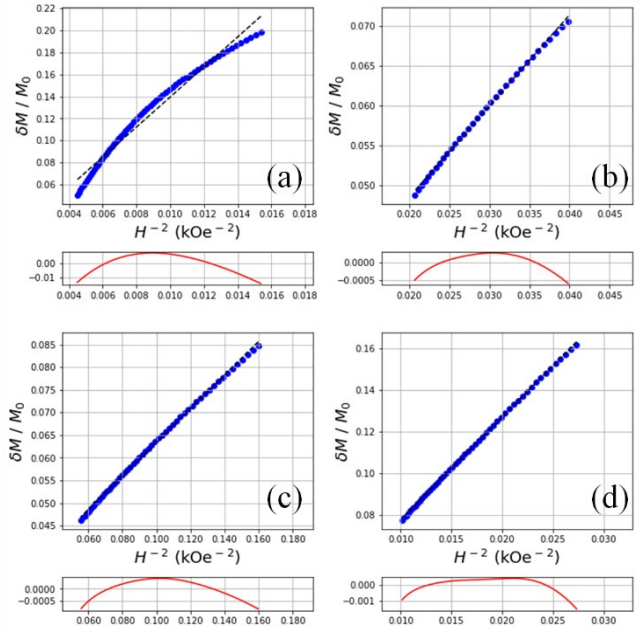

**Figure 6.** Random magnetic behavior fits of δ$M/M_0$ vs $H^{-2}$ for the HF (a), HF-Ni (b), HF-Cu (c), and HF-Mn (d) samples. Below each plot, we represent the error, taken as the difference between the experimental points and the theoretical line.

These results suggest that barium hexaferrite could become a RAM by properly adding divalent cations. In our case, and following the structural characterization previously discussed, the physical basis for this state would be the coexistence, within each particle, of nanocrystallites of the two kinds: the hard magnetic barium hexaferrite and the soft magnetic ferrite based on the divalent cation added. Moreover, bearing in mind that the preparation process should bring about samples with a primary contribution from nanocrystallites of the divalent ferrite, we could deem the nanocrystallites of the original barium hexaferrite as magnetic impurities that would introduce a certain degree of disorder, which would change from one point to another across the samples [50]. In this situation and conforming to the image we suggested after the structural analysis, we could picture the microstructure of these samples as a heterogeneous mixture of two phases, each with its magnetic properties and randomly oriented anisotropy axes.

### 3.3. Microwave properties

It is a well-known fact that composites containing alkaline earth hexaferrite perform as superb microwave absorbers in the GHz frequency range [42], [43], [49]. We have seen before that HF-Cu is the modified sample that experiences the major softening of its magnetic properties compared with the pure hexaferrite. Consequently, the energy associated with its $M(H)$ cycle is insignificant (2.0 kOe emu g$^{-1}$), and, therefore, it shows the most promising properties for high-frequency applications, as stated in the introduction. For this reason, we will focus the microwave analysis on this sample to evaluate the potential of RAMs as microwave absorbers.

We measured the frequency dependence of the complex reflection and transmission $S$-parameters ($S_{11}$ and $S_{21}$, respectively) between 2 and 20 GHz. We ensured the quality of the measurement by verifying the contact between the two terminals once we had placed the sample inside. We did not detect low-frequency losses, the signal was stable along all the frequency ranges, and the signal-to-noise ratio was high. The only remarkable aspect was the presence of a peak at approximately 4 GHz. That was a pure geometrical artifact in the measurement because the specific thickness and refraction index might make the wave resonate at a characteristic frequency, leading to the appearance of this peak. We experimentally validated the peak shifts for different sample thicknesses. Therefore, we cannot consider this peak as an intrinsic absorption of the material. For this reason, we will represent all the results from 7 to 20 GHz to avoid this artifact.

We processed these data using the Nicolson-Ross-Weir method [51], [52], [53] and obtained the complex relative dielectric permittivity, $\hat{\varepsilon}_r$, and the complex relative magnetic permeability, $\hat{\mu}_r$, which appear in Figure 7. The material shows a dominant dielectric behavior over the magnetic one. The low magnetic response is expected at these frequencies, considering that most magnetic materials experiment a rapid magnetic relaxation in the MHz frequency range, which drops the permeability at higher frequencies [54].

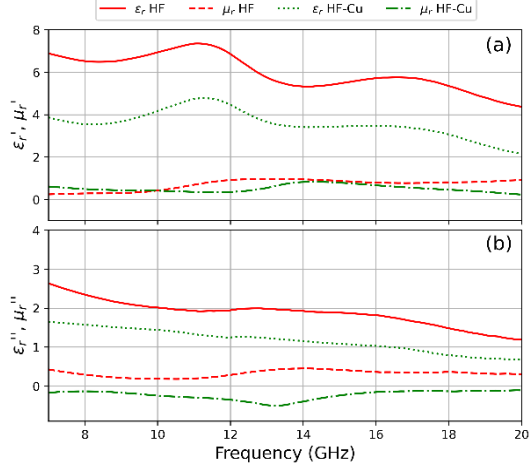

**Figure 7.** Frequency dependence of the complex relative dielectric permittivity ($\hat{\varepsilon}_r$) and complex relative magnetic permeability ($\hat{\mu}_r$) for the HF and HF-Cu samples (see legend for details). Panels (a) and (b) show, respectively, the real and imaginary parts of each one of these magnitudes.

With these magnitudes, we calculated the reflection loss coefficient, $R_L$, according to the principle of equivalent transmission line [51], [52]:

$$R_L(dB) = -20 \log|[(Z/Z_0) - 1]/[(Z/Z_0) + 1]|, \qquad (4)$$

where

$$Z = Z_0\sqrt{\mu_r/\varepsilon_r} \tanh[(j\, 2\pi f t/c)\sqrt{\mu_r \varepsilon_r}] \qquad (5)$$

is the impedance of the sample, $Z_0$ is the impedance of free space, $f$ is the frequency of the microwave, $c$ is the speed of light in vacuum, and $t$ is the thickness of the sample. The fact that the reflection loss coefficient depends on two variables, the microwave frequency and the thickness of the material, allows for a 2D representation in which the color changes with the intensity (in dB) of $R_L$, as it appears in Figure 8. In the first place, the figure reveals that the absorption does not depend linearly on thickness, as stated in recent literature [55]. The comparison of both panels shows that the highest absorption for HF-Cu appears at lower frequencies than in pure HF. Moreover, HF-Cu shows a large plateau of light red (~ 30 dB) absorption [see panel (b)], which remarks the promising capacity of RAMs for absorbing in a broad frequency range.

The typical threshold value to consider a material as a microwave absorber is –10 dB [56], [57], [58]. Thus, the highest values reached by $R_L$ in both samples –for the discrete thicknesses here evaluated– turn to be very competitive compared with values found in pure barium hexaferrite, in modified or composite-like hexaferrite [59], [60], [61], [62], [63], [64], [65] or even in other kinds of materials [66], [67]. In addition, a recent theoretical study [34] postulates that the optimal condition for broadband absorption is having a grain size of about the domain wall thickness in a conventional ferromagnet, which is in the ballpark of 50 nm [68]. That means that in the case of sample HF-Cu, for which the average crystallite size is about 40 nm, as measured from XRD, we have a RAM with an ideal nanostructure for broad microwave absorption. These findings agree well with the recent work that proves that, under actual radar conditions, layered systems containing these nanocomposites can be excellent candidates for shielding applications [55].

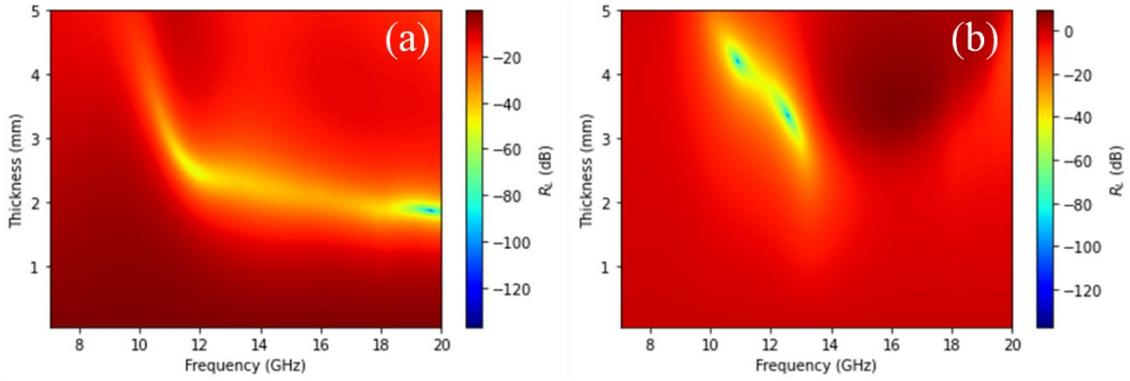

**Figure 8.** 2D representation of the estimated reflection loss coefficient, $R_L$, as a function of the sample thickness (vertical axis) and the microwave frequency (longitudinal axis) for the HF (a) and HF-Cu (b) samples. The color scale goes from approximately 0 dB (dark red) to –120 dB (dark blue).

For most stealth applications where these functional materials are needed, especially in lightweight sectors such as aviation, sub-millimetric coatings with broad absorption bandwidths are ideal candidates [56], [69]. Even though the results do not seem to manifest substantial differences between the two samples regarding absolute absorption, Figure 8 suggests that the HF-Cu sample presents broader absorption regions than HF. To avoid extracting conclusions from qualitative considerations, we numerically studied the absorption bandwidth for the two samples. To do so, we measured the frequency range at which the absorption peaks present an $R_L$ value equal to or better than a determined threshold value. Given that the selection of the $R_L$ threshold affects the results and depends on the particular application devised for the nanocomposite we are working with, we computed the bandwidth along a coherent range of $R_L$ thresholds.

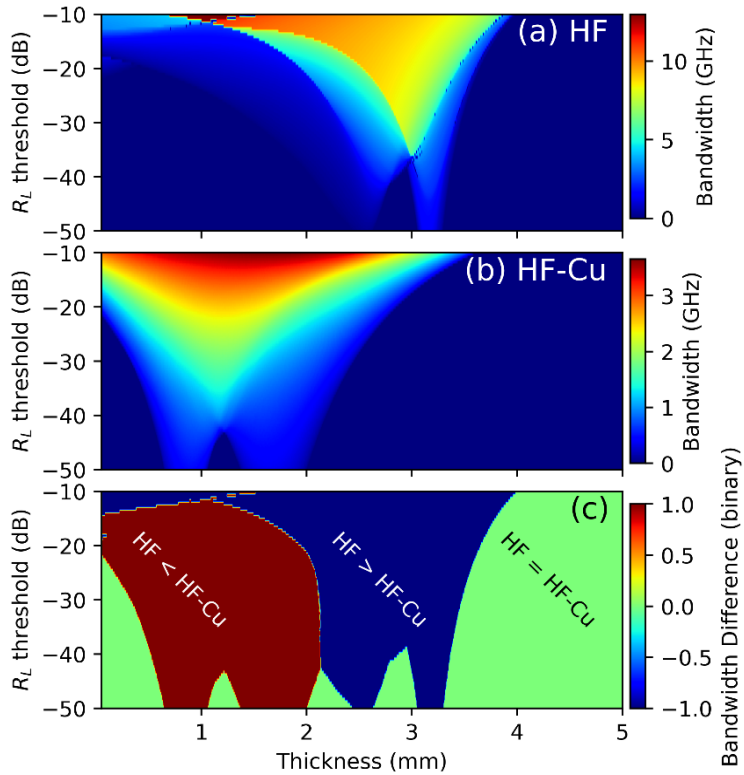

**Figure 9.** Analysis of the absorption bandwidth for the two samples. Panels (a) and (b) show the dependence of the absorption bandwidth on sample thickness and $R_L$ threshold for sample HF and HF-Cu, respectively. Panel (c) represents a binary plot with the difference between the signals presented in panels (b) and (a). The positive regions

identify conditions in which the absorption bandwidth of the HF-Cu sample is larger than the one for HF. The negative results identify the opposite scenario.

After a first look at panels (a) and (b) in Figure 9, one may observe that both samples have a range of thicknesses where the peaks are within the measured frequency range. When the thickness is too small (especially in the case of HF) or too large, the intensity of the absorptions drops quickly. Even though HF may lead to wider absorptions than HF-Cu, it requires thicker samples to reach an optimum response. The higher the expected $R_L$ is, the narrower the absorption bandwidth is, and fewer are the adequate thicknesses that lead to optimum results. Panel (c) allows for a direct comparison between the two samples along the complete experimental domain. This plot helps to identify quickly the conditions where each material yields a broader absorption. To do so, we computed the difference between the absorption bandwidth at each coordinate and represented it in a binary mode: +1 regions (dark red) are those where HF-Cu has a broader absorption peak than HF, while –1 regions (dark blue) define the opposite scenario. Neutral (green) results appear when neither sample has significant absorption peaks. This panel indicates that the bandwidth of HF-Cu is broader when significant absorption levels ($|R_L| > 15$ dB) are required in thin (< 2 mm) samples. Overall, the results reported here prove the potential of random anisotropy magnets to improve broadband absorption capacities in thin systems.

The magnetic properties, together with the results obtained for the microwave measurements, can be understood in the framework of the recent theoretical works developed by Garanin and Chudnovsky [30], [31], [32], [33], [34], devoted to the study of the absorption of microwaves by RAMs. Even though their microscopic model was originally elaborated for magnetic systems at the atomic scale, as it focused on the interaction of spins with a microwave field in a RAM, in their most recent papers [33], [34] the authors enlarge the picture to include polycrystalline systems consisting of nanoscale ferromagnetic crystallites, or systems sintered from nanoscale ferromagnetic grains. According to this model, high microwave absorption occurs in a broad GHz frequency range due to the random distribution of ferromagnetic regions (Imry-Ma domains) on sizes and effective anisotropy. At a fixed frequency, the microwave field makes the spins or magnetic moments oscillate in resonance inside isolated regions, and the power absorbed depends on the number of spins or magnetic moments involved. In this context, having discussed the random nature of the magnetic properties of our modified hexaferrites and considering the estimates we have found for the reflection loss coefficient, we believe these samples fit in with this theoretical model for microwave absorption by RAMs.

## 4. Conclusions

In this work, we have experimentally proved that adding divalent cations into a barium hexaferrite forms a second ferrite structure instead of dissolving directly into the hexaferrite, giving rise to a magnetic nanocomposite. We have observed that nanoparticles agglomerate during thermal treatment and form micrometric clusters. From a magnetic point of view, we have detected a change in the magnetic properties of the barium hexaferrite and a transition to a random magnetic behavior when we add the divalent cations. We understand our materials as a system of nanometric independent grains with randomly oriented anisotropy axes. That magnetic state leads to a broad absorption in the GHz frequency range, which agrees well with recent theoretical models that postulate random magnets as excellent microwave absorbers.

The results reported here may be of great technological interest as they open the possibility of preparing ceramic random anisotropy magnets, which should play a crucial role in the industry of high-frequency components made of ferrite materials instead of rare-earth-based intermetallic compounds. Such materials have the advantage of being inexpensive, easy to prepare, and chemically stable. Microwave characterization has also shown that these materials have a strong potential for radar and stealth technology. Given their powdered nature, the synthesized materials could easily integrate into coatings for shielding objects or vehicles. Furthermore, we have proved that these modified nanocomposites widen the absorption peaks for thin (< 2 mm) layers, and this may have a direct positive effect on lightweight radar absorption applications.

## 5. Acknowledgments

The U.S. Air Force Office of Scientific Research (AFOSR) [grant number FA8655-22-1-7049] supported this work. Joan Manel Hernàndez and Antoni García-Santiago thank the Departament de Recerca i Universitats de la Generalitat de Catalunya [grant number 2021SGR00328].

## 6. Author declarations

### 6.1. Conflict of interest

The authors have no conflicts to disclose.

### 6.2. Author contributions

Jaume Calvo-de la Rosa: Conceptualization, Data curation, Formal analysis, Investigation, Methodology, Writing – original draft, Writing – review & editing

Antoni García-Santiago: Conceptualization, Formal analysis, Investigation, Methodology, Writing – original draft, Writing – review & editing

Joan Manel Hernàndez: Funding acquisition, Methodology, Supervision

Marc Vazquez-Aige: Data curation, Investigation

Jose Maria Lopez-Villegas: Methodology, Resources, Supervision, Validation, Writing – review & editing

Javier Tejada: Conceptualization, Funding acquisition, Resources, Supervision, Validation, Writing – review & editing

### 6.3. Data availability

The data supporting the findings of this study are available from the corresponding author upon reasonable request.